\documentclass{aa}  
\usepackage{soul}
\usepackage{graphicx}
\usepackage{float}
\usepackage{txfonts}
\usepackage[colorlinks=true,linkcolor=blue,citecolor=blue, filecolor=blue, urlcolor=blue]{hyperref}
\usepackage{amsmath}
\usepackage{amssymb}

\usepackage[table]{xcolor}
\usepackage{natbib}
\bibpunct{(}{)}{;}{a}{}{,}
\usepackage[normalem]{ulem}
\usepackage{mathrsfs}
\usepackage{calligra}
\usepackage{threeparttable}
\makeatletter
\let\linenumbers\relax

\makeatother

\begin{document} 

\titlerunning{Impact of line-of-sight structure on weak lensing observables of galaxy clusters}

\title{Impact of line-of-sight structure on weak lensing observables of galaxy clusters}

\author{Felix Vecchi \inst{1}\thanks{felix.vecchi@epfl.ch}
          \and David Harvey\inst{1} 
          \and James Nightingale\inst{2} 
          \and Matthieu Schaller \inst{3, 4}
          \and Joop Schaye \inst{4}
          \and Ethan Tregidga \inst{1}
          }

\institute{
Laboratoire d’Astrophysique, EPFL, Observatoire de Sauverny, 1290 Versoix, Switzerland
\and School of Mathematics, Statistics and Physics, Newcastle University, Herschel Building, Newcastle-upon-Tyne, NE1 7RU, UK
\and Lorentz Institute for Theoretical Physics, Leiden University, PO Box 9506, NL-2300 RA Leiden, The Netherlands
\and Leiden Observatory, Leiden University, PO Box 9513, NL-2300 RA Leiden, The Netherlands
}

\abstract{
    Weak gravitational lensing observations of galaxy clusters are sensitive to all mass along the line-of-sight, introducing systematic and additional statistical uncertainties due to intervening structures. We quantify the impact of these structures on the recovery of mass density profile parameters using 967 clusters from the highest-resolution FLAMINGO simulation. We construct mock weak lensing maps, which include both single source plane mocks at redshifts up to $z_\text{s}\leq3$, and Euclid-like mocks with a realistic source redshift distribution. Applying Bayesian inference with \texttt{Nautilus}, we fit spherical and elliptical Navarro-Frenk-White models to recover the cluster mass, concentration, axis ratio, and centre, which we use to measure the brightest cluster galaxy (BCG) offset from the potential centre, or ``BCG wobble''.
    We find that the spherical model fits clusters along under-dense sight-lines better than those along over-dense ones. This introduces a positive skew in the relative error distributions for mass and concentration, which increases with source redshift. In Euclid-like mocks, this results in a mean mass bias of $+5.3\pm1.4$\% (significant at $3.5\sigma$) when assuming a spherical NFW model. We also detect a mean axis ratio bias of $-2.0\pm0.7$\% ($2.9\sigma$), with no significant bias in concentration. We measure a BCG wobble of $\sim 14$ kpc in our Euclid-like mocks, with negligible contribution from line-of-sight structure. Furthermore, we predict the scatter in mass estimates from future weak lensing surveys that have mean source redshifts $z_\text s \gtrsim 1.2$ (such as the Nancy Grace Roman Space Telescope), will be dominated by line-of-sight structure and hence assuming a diagonal covariance matrix will lead to an overestimation of the precision. We conclude that cluster weak lensing pipelines should be calibrated on simulations with lightcone data in order to properly account for the significant impact of line-of-sight structure.
}

\keywords{Gravitational lensing: weak; Galaxies: clusters: general}

\maketitle

\section{Introduction}

Clusters of galaxies, hereafter referred to as clusters, are the most massive collapsed systems in the Universe and represent the end products of hierarchical structure formation. These massive structures have been built up through cosmic history, making them sensitive to the cosmological parameters that govern our Universe.
For example, cluster counts and masses can be used to trace the high-end of the halo mass function, which is dependent on cosmological parameters (e.g. \citealp{Allen_2007}). In addition, the (standardised) fraction of gas in clusters depends on cosmological distances, making it cosmology dependent (e.g. \citealp{Ettori_2009}). 

Being dark matter dominated systems, the three dimensional mass density profile of clusters is typically well described by the Navarro-Frenk-White (NFW) profile (e.g. \citealp{Rines_2013, Niikura_2015, Child_2018}), a broken power law with a transition at a characteristic ``scale radius''  \citep{Navarro_1996, Navarro_1997}. The ratio of the virial radius to the scale radius is referred to as the concentration, which, although subject to significant scatter, is related to the cluster's mass and redshift (e.g. \citealp{Zhao_2003}). This concentration-mass-redshift relation is sensitive to cosmology and can therefore be used to constrain cosmological parameters (e.g. \citealp{Correa2015c, Ludlow_2016, L_pez_Cano_2022}). 

Clusters of galaxies are found in the densest nodes of the cosmic web, making them ideal laboratories for probing the nature of dark matter. Specifically, their internal structure provides a means to constrain the particle properties of dark matter, as any subtle modification that alters its dynamics will be amplified in these environments.
For example, a finite dark matter self-interaction cross-section leads to cored inner regions of galaxy clusters. In the event of a major merger, the brightest cluster galaxy (BCG) — the large central elliptical galaxy typically found in clusters — can become offset from the bottom of the underlying gravitational potential, which is dominated by dark matter. This offset can persist long after virialisation via a ``BCG Wobble''. The amplitude of this wobble scales with cross-section and provides a viable way to probe the dark matter self-interaction cross-section \citep{Kim_2017, Harvey_2017, Harvey_2019}.

Additionally, dark matter self-interactions cause the dark matter halo of a cluster to become more spherical. This effect has been used to constrain the dark matter self-interaction cross-section \citep{Miralda_Escude_2002}, although the observational feasibility of these measurements is still uncertain \citep{Harvey_2021, Robertson_2023}. Furthermore, the shapes and orientations of clusters are correlated with feedback processes and star formation (e.g. \citealp{Bryan_2013,Velliscig_2015,Donahue_2016}) and can therefore help understand these processes.

Using galaxy clusters for cosmological purposes requires reliable characterisation of their mass density profile, including their total mass, concentration, shape, and centre. The density profile can be constrained using dynamical tracers, such as the X-ray emission of the intracluster gas (see e.g. \citealp{Ettori_2013} for a review). However, such methods often rely on assumptions about the cluster's dynamical state, which can introduce biases (e.g. \citealp{Eckert_2016}). 

The strongly warped spacetime around clusters distorts the images of background source galaxies, causing them to act as gravitational lenses. This distortion is sensitive to the second-order derivative of the projected gravitational potential, and therefore provides a direct way to probe the total projected mass density profile, independent of assumptions about the cluster's dynamical state. The efficiency of gravitational lensing increases with the distance to the source galaxies and is maximal for lenses at a distance slightly closer than halfway between the observer and source.
Gravitational lensing is subdivided into two distinct regimes. In the strong lensing regime, relevant to the inner regions of clusters, background galaxies are highly distorted, forming giant arcs or even multiple images. Beyond these central regions, the weak lensing regime applies, where distortions are more subtle and can only be extracted statistically from many source galaxies. In this work, we focus on weak gravitational lensing, as this can be readily applied to study large cluster samples.

Like all observational techniques, constraining the mass density profile of clusters via weak gravitational lensing observations is subject to various systematic uncertainties. These systematics can be broadly classified into three main categories:
\begin{enumerate}
    \item Systematic errors associated with the background source galaxies: These include additive and multiplicative  shape measurements bias and photometric redshifts errors, which have all been studied extensively (e.g. \citealp{Hoekstra_2008, Refregier_2012, Mandelbaum_2014, Hoekstra_2015, Bernstein_2010, Varga_2019}).
    \item Systematic errors associated with the modelling of the cluster's mass density profile: Clusters are complex objects with non-trivial formation histories, and hence models often fail to fully capture their structure. Moreover, baryons in the cluster are subject to cooling and various feedback processes and therefore alter the total mass density profile of clusters. Typically, this is not taken into account in the modelling and, therefore, these ``baryonic effects'' pose a systematic uncertainty (e.g. \citealp{Debackere_2021, Grandis_2021, giocoli2025}). Furthermore, dark matter halos are triaxial in nature (e.g. \citealp{Allgood_2006}), and their orientation with respect to the line-of-sight, which cannot be measured from the projected lensing signal, leads to a systematic uncertainty (e.g.  \citealp{Bahe_2012, Lee_2018, giocoli_2024}). 
    \item Systematic uncertainties arising from other structure along the line-of-sight: Given the width of the lensing kernel, other structure along the line-of-sight also contributes to the lensing signal, introducing an additional source of systematic uncertainty. Nearby structures, such as those within the same filament as the cluster, are considered 'correlated', as, for example, their major axis tends to align with the filament \citep{Kasun_2005}. In contrast, more distant structures, part of the large-scale structure, are considered 'uncorrelated', with their contribution being independent of the cluster's properties. This last category will be the focus of this work, although systematics associated with mis-modelling are also naturally incorporated into our analysis.

\end{enumerate}

Previous studies on the impact of line-of-sight structure on weak lensing observables focussed mainly on the cluster's mass and relied on analytical models, dark-matter-only simulations, or a combination of both.
One of the first studies examined a small sample of clusters in an N-body simulation with an integration length of $256$ Mpc$/h$, including correlated structures and part of the uncorrelated structure \citep{Metzler_2001}. Their analysis, based on aperture mass densitometry (or the $\zeta$-statistic) as a weak lensing mass estimator, found that line-of-sight structure increased the scatter in mass estimates and introduced a positive bias. A later study by \cite{Wu_2006} demonstrated that performing this analysis with shear data for a larger sample of clusters mitigated this bias while preserving the increased scatter.

In parallel, \citealp{Hoekstra_2001, Hoekstra_2003} employed an analytical approach to quantify the influence of uncorrelated structures along the full line-of-sight on the inferred cluster mass and concentration under the assumption of an NFW profile. These theoretical predictions were later validated by \cite{Hoekstra_2011}, who perturbed the shear profile of a spherically symmetric NFW halo with random sight-lines through the Millennium dark-matter-only simulation \citep{Springel_2005}, using a self-consistent source redshift distribution. Their results confirmed that averaging over a sufficiently large number of sight-lines with over- and under-densities leads to increased scatter in weak lensing-derived mass and concentration estimates without introducing a significant bias.
Subsequent studies reinforced these conclusions. For example, \cite{Becker_2011} investigated the role of correlated structure on weak lensing observables, testing integration lengths ranging from 6 to 400 Mpc, with a fixed source redshift of $z_\text{s}=1$, further demonstrating that line-of-sight structure increases the scatter in weak lensing mass estimates.

This study revisits the impact of line-of-sight structure on weak lensing observables of galaxy clusters. Using a large hydrodynamical simulation from the FLAMINGO project \citep{Schaye_2023,kugel2023}, we improve on previous studies by self-consistently forward-modelling both the cluster and structure along the full line-of-sight while incorporating baryonic effects.
We analyse cluster mass, concentration, axis ratio, and the BCG wobble, assessing the relative significance of line-of-sight structure compared to shape noise — the primary source of statistical error in weak lensing. Specifically, we vary the source plane redshift to highlight a fundamental trade-off in gravitational lensing: while increasing the source redshift increases the lensing efficiency, it also extends the line-of-sight, thereby incorporating a greater number of intervening structures. Moreover, we incorporate a realistic source redshift distribution, to quantify the impact of line-of-sight structure on cluster weak lensing studies with upcoming Euclid data \citep{laureijs2011,Scaramella_2022}.
Finally, we compare the true scatter in the inferred parameters to the Bayesian-estimated uncertainties to evaluate whether the impact of line-of-sight structure is appropriately accounted for within the error budget.

This paper is organised as follows: Section \ref{sec:simulation} describes the FLAMINGO simulation run used in this study, the selection of our simulated cluster sample and the methods used to extract mass maps. Section \ref{sec:lensing_theory} outlines the theoretical framework of weak gravitational lensing and describes the setup of our different mock weak lensing maps. In Section \ref{sec:fit}, we detail the modelling choices for the cluster mass density profile. Section \ref{sec:results} presents and discusses the results of our analysis. Finally, we conclude in Section \ref{sec:conclusions}. In this work, bold symbols denote vector quantities.

\section{Simulation data}
\label{sec:simulation}
\subsection{FLAMINGO}
In this project, we work with the FLAMINGO suite of cosmological hydrodynamical simulations, which offer an unprecedented combination of box size and resolution, designed specifically for studying galaxy clusters and large-scale structure \citep{Schaye_2023, kugel2023}. These simulations were run with \texttt{SWIFT} \citep{Schaller_2024} and calibrated on the stellar mass function at $z=0$ and the cluster gas fraction at low-redshift \citep{kugel2023}. FLAMINGO has been demonstrated to accurately reproduce X-ray observations of clusters, including their temperature, density, and entropy profiles \citep{braspenning2024}.

The FLAMINGO simulations' large volume, accompanied by lightcone output, enables us to generate mock weak lensing maps that include structure along the line-of-sight for a large sample of clusters. For this work, we use the high-resolution fiducial hydrodynamical run, as this allows us to compute the high-resolution mass maps required for our analysis. This high-resolution run has a total of $10^{11}$ particles ($3600^3$ baryon and dark matter particles each plus 2000$^3$ neutrino particles), initial baryonic and cold dark matter particle masses of $m_{\text{gas}}=1.34\times 10^8 \text{M}_\odot$ and $m_{\text{dm}}=7.06\times 10^8 \text{M}_\odot$ respectively, and a fixed physical gravitational softening length of $2.85 \text{ kpc}$ below $z=2.91$. This simulation run assumes the Dark Energy Survey Year 3 \citep{Abbott_2022} cosmology (3x2pt plus external constraints), assuming a spatially flat Universe with a neutrino mass of $\Sigma m_\nu c^2 = 0.06$ eV. To ensure consistency with the simulation, we adopt the same cosmology throughout this paper. For our analysis, the relevant parameters are: $H_0 = 68.1$ km/s/Mpc and $\Omega_{\text{m},0} = 0.306$.

The simulation includes lightcone output of particle data extending to $z=0.25$. Additionally, full-sky HEALPix maps of all matter components are generated in redshift shells of $\Delta z=0.05$ up to $z=3$. Structure finding was performed on simulation snapshots (spaced by $\Delta z=0.05$) using HBT-HERONS (Hierarchical Bound Tracing - Hydro-Enabled Retrieval of Objects in Numerical Simulations; \citealp{Han_2017, moreno2025}). HBT-HERONS detects central halos and tracks their evolution across snapshots, including merger events. The most bound black hole particle serves as a marker for tracking the halo through the lightcone. This enables halos to be matched and stored in lightcone-halo files, which we use for our sample selection.
For the structures identified by HBT-HERONS, a wide range of additional properties are computed using SOAP \citep{McGibbon_2025}, a tool specifically designed for the FLAMINGO project.

\subsection{Sample selection}
We construct a sample of clusters from the SOAP-HBT-HERONS catalogues that accompany the FLAMINGO suite of simulations. To ensure that the weak lensing signal-to-noise is high enough for all clusters in the sample, we only include halos with $M_{200}\,>3\times 10^{14} \text{ M}_\odot$. Where $M_{200}$ is the mass enclosed by a spherical aperture with radius $R_{200}$ which is the radius at which the average density inside the aperture becomes 200 times the critical density of the Universe. In our study we want to rule out any influence of lens redshift dependent quantities, such as the field-of-view or source galaxy number density in physical length units. Therefore we restrict the sample to a thin redshift slice. For the source redshift distribution expected for Euclid \citep{euclid1}, the lensing efficiency (see Equation \ref{eq:zeff}) peaks at $z_\text{l}=0.23$. Therefore, we choose to select only clusters with $0.20 < z < 0.25$ for our sample. Lastly, we ensure that each cluster in our full sample has a unique HBT-HERONS track ID, guaranteeing that all clusters are completely independent of each other. Imposing these conditions results in a final sample size of 967 clusters.

Generating lightcone data for lookback distances larger than the box length requires tiling of the simulation box to represent a larger volume. These ``box replications'' can result in the same structures appearing multiple times at different redshifts. 
We investigated whether box replications affect our results by masking out clusters positioned within $\pm$ 9 Mpc along the main axes of replication (see e.g. \citealt{chen2024}).
This distance corresponds roughly to twice the size of our physical field-of-view at $z\approx 1.5$, where the angular diameter distance peaks. We found that masking these clusters (56) did not significantly alter our conclusions, so we do not exclude these clusters from our analysis. 

\subsection{Mass maps}
\label{sec:massmaps}
We generate mass maps for the selected clusters with a resolution of 2 arcsec, which corresponds to a physical size of 7.4 kpc at $z=0.225$ — the centre of the redshift range of our sample. This makes our pixel resolution roughly equal to 60\% of the \cite{Ludlow_2019} criterion for numerical convergence of $r_\text{conv}=0.055l$, with $l$ being the mean comoving inter-particle separation. We aim to probe the cluster's weak lensing signature to well beyond their virial radius. Therefore, we adopt a field-of-view of 18.0 arcmin, corresponding to a physical size of 4.0 Mpc at $z=0.225$, ensuring that the virial radius of the most massive clusters is fully contained within the map. The field-of-view and resolution define a pixelized grid onto which we create our mass maps. 

For each cluster, we divide the line-of-sight into shells out to $z=3$ and compute the corresponding mass map for each shell. These shells are used in the next section as input for generating mock weak lensing maps. Following the structure of the HEALPix maps, we set the shell edges at intervals of $\Delta z = 0.05$. 
For a given shell with $z < 0.25$ we generate the total mass map from the particle lightcone output by summing the masses of all particles (cold dark matter, gas, stellar, black hole, neutrino) falling into each pixel on our grid. 
For $z>0.25$ the lightcone output is provided as full-sky HEALPix maps, with no lightcone particle data stored. To generate mass maps for these redshifts, we up-sample the 13 arcsec HEALPix maps to the required 2 arcsec resolution. We up-sample by linearly interpolating between the pixels, which introduces a smoothing effect on line-of-sight structure for $z>0.25$. 
An example of this is provided in the in Figure \ref{app:LOS-convergence-map} of Appendix \ref{appendix:smoothing}. We have verified that the smoothing of line-of-sight structure does not impact our final results. For details on this test, we refer the reader to Appendix \ref{appendix:smoothing}.

\begin{figure*}[ht]
    \includegraphics[width=\textwidth]{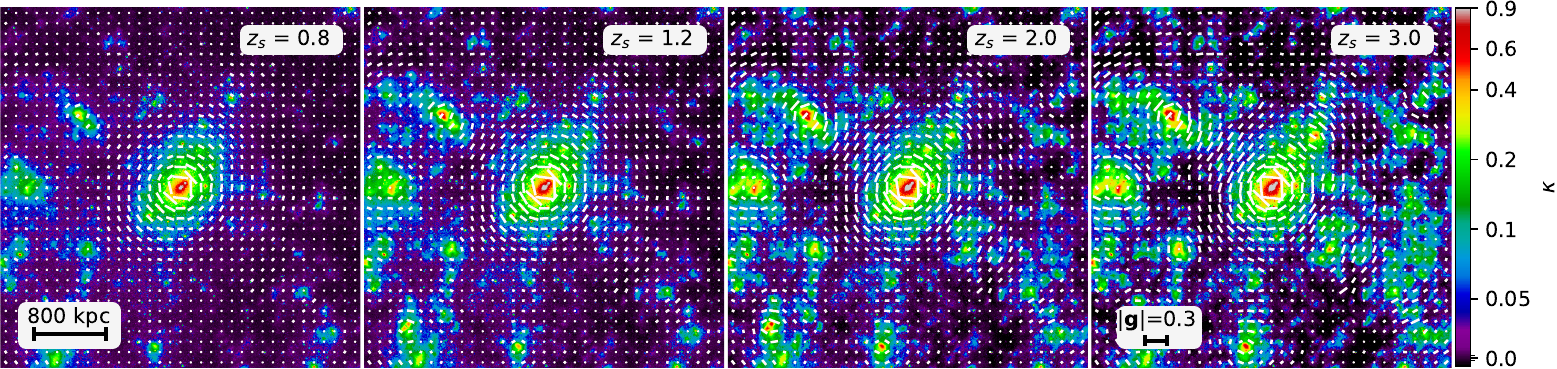}
    \caption{High resolution convergence maps overlaid with reduced shear maps of an example cluster at $z=0.231$ with $M_{200}=9.9\times 10^{14} \text{ M}_\odot$. We show different lensing maps from left to right for source plane redshifts of $z_\text{s}$ = 0.8, 1.2, 2.0 and 3.0. For legibility, the reduced shear map has been down-sampled by a factor 30. At higher source redshifts, the gravitational lensing efficiency increases, but so does the amount of line-of-sight structure.}
    \label{fig:LOSS_maps}
\end{figure*}

\section{Weak gravitational lensing}
\label{sec:lensing_theory}

\subsection{Weak lensing theory}
Gravitational lensing is the phenomenon where the image of a background source is distorted by the presence of a foreground  gravitational field. In this work, the gravitational field belongs to a cluster of galaxies (with perturbations belonging to structure along the line-of-sight). 
In this section, we introduce the relevant quantities for this study and refer the reader to \cite{Umetsu_2020} for details on cluster-galaxy weak lensing.

In this work, we construct weak lensing maps that contain structure along the full line-of-sight. To this end, we work within the thin lens approximation, defining a series of thin lenses between the observer and the source plane, corresponding to the shells described in section \ref{sec:massmaps}. After calculating the convergence in each shell, we apply the Born approximation, summing these contributions to generate a convergence map that includes all line-of-sight structure up to an arbitrary source plane redshift below $z=3$. We do not perform full multi-plane ray tracing, as this would be computationally prohibitive. Furthermore, the differences between the Born approximation and full ray tracing are minimal, at least on the level of the weak lensing angular power spectrum \citep{ferlito2025, Broxterman_2025}.

Following \cite{Umetsu_2020}, the convergence for a given shell is
\begin{equation}
    \kappa_\text{shell} = \int_{\chi(z_\text{min})}^ {\chi(z_\text{max})} ad\chi\cdot (\rho - \bar{\rho})\cdot \Bigg(\frac{c^2}{4\pi G}\frac{D_\text{s}}{D_\text{l} D_\text{ls}}\Bigg)^{-1},
\end{equation}
where $\rho$ is the matter density along the line-of-sight and $\bar{\rho}$ is the mean matter density of the Universe at the redshift of the lens. The integral is calculated over the comoving distance $\chi$ at the minimum redshift of the shell to the comoving distance at the maximum redshift of the shell. We assume that the redshift bins of our shells are small enough such that the matter in each shell can be approximated as collapsed into a single thin lens at $z=z_\text{l}$. Then
\begin{equation}
    \kappa_\text{shell}(\Vec{\theta}) = \frac{\Sigma_\text{shell}(\Vec{\theta})}{\Sigma_\text{crit}},
\end{equation}
where we defined the critical surface density as
\begin{equation}
    \Sigma_\text{crit}(z_\text{l},z_{\text{s}}) = \frac{c^2}{4\pi G}\frac{D_\text{s}}{D_\text{l} D_\text{ls}}.
\end{equation}
The inverse of this quantity is proportional to the lensing efficiency (see Equation \ref{eq:G} below), which increases as a function of source redshift, and will be used throughout this paper.
The surface over-density in a given shell is calculated as
\begin{equation}
    \Sigma_\text{shell}(\Vec{\theta}) = \frac{M_\text{pix}}{A_\text{pix}(z_\text{l})} - \Sigma_\text{mean}(z_\text{l}) ,
\end{equation}
where $M_\text{pix}$ is the mass in a pixel of our mass map, $A_\text{pix}(z)$ the physical area of the pixel and $\Sigma_\text{mean}$ is the mean surface mass density at redshift $z_\text{l}$,
\begin{equation}
   \Sigma_\text{mean}(z_\text{l}) = \rho_\text{c,0}\cdot\Omega_\text{m,0}\cdot(1+z_\text{l})^3 \cdot \Delta x_\text{shell},
\end{equation}
where, $\rho_\text{c,0}$ is the critical density of the Universe at $z=0$ and $\Delta x_\text{shell}$ is the width of the shell in physical units. 
In this way, we calculate the convergence map for each shell, assuming the central redshift of the bin as the lens redshift. For the shell containing the cluster, we use the cluster's redshift as the lens redshift.

Both shear and convergence are second-order derivatives of the effective lensing potential. This means that they are related to each other in Fourier space. We apply the Kaiser-Squires method \citep{Kaiser_93}, which leverages this property to transform our convergence maps into a shear maps.

To minimise boundary effects of the Fourier transform, we zero-pad the input convergence map to 5 times the original size. We also tested ``true padding'', where the mass map of the shell is extended to five times its original size. However, this yielded results identical to zero padding while being significantly more computationally expensive. 

After obtaining both convergence and shear for all the shells between the observer and the source plane, we sum all the contributions to construct the final convergence map ($\kappa$) and final shear map ($\vec{\gamma}$), which accounts for structure along the full line-of-sight.

Then, we can simply compute the reduced shear $\vec{g}$ via
\begin{equation}
 \vec{g} =\frac{\vec{\gamma} }{1-\kappa}.
\label{eq:g}
\end{equation}
In a weak lensing analysis, it is often assumed that $\vec{g}\simeq\vec{\gamma}$. However, this assumption breaks down in clusters, so we explicitly model $\vec{g}$.

We show the convergence and overlaid reduced shear maps of an example cluster for different source plane redshifts in Figure \ref{fig:LOSS_maps}. As we increase the source plane redshift, two important effects can be seen. First, the appearance of increasing amounts of line-of-sight structure and second, an increase in the lensing efficiency.

In a realistic cluster weak lensing analysis, source galaxy redshifts are distributed over a broad range, with the exact shape of the distribution depending on the specifics of the survey. This is captured in the source redshift distribution, $n(z_\text{s})$, which we define such that $\int n(z_\text s) d z_\text s=1$. Given $n(z_\text{s})$ and the lens redshift, the lensing efficiency can be computed with
\begin{equation}
    G(z_\text l) = \int_{z_\text l}^{\infty} n(z_\text s) \frac{D_\text{ls}(z_\text l,z_\text s)}{D_{\text s}(z_\text s)}dz_\text s.
    \label{eq:G}
\end{equation}

In weak lensing studies of clusters, the redshifts of the individual source galaxies are often not known. Therefore, it is common practice to adopt a single source plane approximation, in which all source galaxies are assumed to lie on a thin plane at an effective redshift, $z_\text{eff}$. For low-redshift ($z_\text l\lesssim0.3$) lenses, this effective source redshift is obtained by solving

\begin{equation}
    \frac{D_\text{ls}(z_\text l,z_\text{eff})}{D_\text{s}(z_\text{eff})} = G(z_\text l).
    \label{eq:zeff}
\end{equation}

This ensures that the lensing efficiency of the approximated single source plane is equal to that of the real distribution of source redshifts.

\subsection{Mock weak lensing maps}
\label{sec: mocks}
In weak lensing, the dominant source of statistical noise is the intrinsic shape of the source galaxy. Following \cite{Bartelmann_2001}, the observed ellipticity\footnote{Where we use the convention that $ |\vec{\epsilon}| = (a-b)/(a+b)$, where a and b are the major and minor axes of the ellipse, respectively.}, $\vec{\epsilon}$ for $|\vec{g}|<1$ is calculated as 
\begin{equation}
    \,\,\,\,\,\,\, \vec{\epsilon} =  \frac{\vec{\epsilon}_\text{int} + \vec{g}}{1 + \vec{\epsilon_\text{int}}\cdot \vec{g}} .
    \label{eq:ellipticity}
\end{equation}
The intrinsic shape distribution of galaxies is well known, with a measured dispersion per component of $\sigma_\epsilon=0.26$, hereafter referred to as ``shape noise". This value is derived from observations in the COSMOS \citep{Leauthaud_2007} and CANDELS \citep{Schrabback_2017} fields using the Hubble Space Telescope, with photometric properties similar to the Euclid VIS instrument. As a result, this value is widely adopted for modelling purposes within the Euclid Collaboration (e.g., \citealp{Martinet_2019, Ajani_2023, Ingoglia_2025}) and will be used for this work.
 
Statistical uncertainty can be reduced by averaging over many galaxies in a given path of sky, however, this approach is ultimately limited by the number density of source galaxies.
In this work, we assume a source galaxy number density of 30 galaxies per square arcminute, consistent with the expected depth of the Euclid survey \citep{euclid1}. For our assumed field-of-view, this corresponds to a total of $98^2$ source galaxies.

To incorporate this, we down-sample our high-resolution reduced shear maps using linear interpolation, placing the source galaxies on a uniform grid.

\begin{table}[t]
    \caption{Overview mock weak lensing maps}
    \label{tab:data}
    \centering
    
    \begin{tabular}{lc c c c}
    \hline\hline
        Designator                & line-of-sight & $\sigma_\epsilon$ & $n(z_\text s)$\\ \hline 
        CL                        & 10 Mpc& 0 & $\delta(z_\text s - \Tilde{z}_\text s)$\\
        CL+LoS                    & < $z_\text{s}$& 0 & $\delta(z_\text s - \Tilde{z}_\text s)$\\  
        CL+ $\sigma_{\epsilon}$   & 10 Mpc& 0.26 & $\delta(z_\text s - \Tilde{z}_\text s)$ \\ 
        CL+LoS + $\sigma_\epsilon$& < $z_\text{s}$& 0.26 & $\delta(z_\text s - \Tilde{z}_\text s)$ \\ 
        Euclid+CL                 & 10 Mpc& 0.26 & $n_\text{euclid}(z_\text s)$ \\ 
        Euclid+CL+LoS             & < $z_\text{s}$ & 0.26 & $n_\text{euclid}(z_\text s)$ \\ 
        \hline 
    \end{tabular}
    \tablefoot{The six different mocks used in this work. First, we have four single source plane mocks for various source plane redshifts $z_\text s=\tilde{z}_\text s$, indicated by the Kronecker delta $\delta$ in the last column: i) ``CL'', isolates the intrinsic shortcomings of the modelling of the cluster's mass density profile; ii) ``CL+LoS'', isolates the effects of structure along the full line-of-sight; iii) ``CL+$\sigma_\epsilon$'', isolates the effects of realistic shape noise; and iv) ``CL+LoS+$\sigma_\epsilon$'', includes the combined effects of the full line-of-sight and realistic shape noise. Then, we have two Euclid-like mocks (with a realistic source redshift distribution, $n_\text{euclid}$) ``Euclid+CL'' and ``Euclid+CL+LoS'', which include shape noise and differ in the inclusion of line-of-sight structure.}
\end{table}

We construct four types of mock weak lensing maps to disentangle the effects of: i) shortcomings of the assumed model; ii) structure along the line-of-sight; and iii) statistical noise arising from the intrinsic shape of source galaxies. To study the dependence of source redshift on these effects, we assume single source planes at redshifts of 0.4, 0.8 ,1.2, 1.6, 2.0, 2.4, and 3.0.
For these source plane redshifts, we define the following mocks:
\begin{itemize}
    \item CL: This mock isolates the intrinsic shortcomings of the modelling of the cluster's mass density profile. It allows us to measure the intrinsic bias or scatter in a given model parameter. We do not add shape noise ($\sigma_\epsilon=0$) and use only a thin line-of-sight shell of $\pm 5$ Mpc around the cluster, which excludes nearly all line-of-sight structure. However, the width of this shell is still larger than the cluster itself and therefore includes part of the correlated line-of-sight structure and its associated systematic uncertainties (see, e.g., \citealp{Bahe_2012}).
    
    \item CL+LoS: This mock isolates the systematic uncertainty introduced by line-of-sight structure. It includes all structure along the line of sight from the observer out to the source plane. No shape noise is included ($\sigma_\epsilon=0$).
    
    \item CL+${\sigma_\epsilon}$: This mock isolates the statistical uncertainty due to shape noise. It includes only a thin line-of-sight shell of $\pm 5$ Mpc around the cluster, while assuming shape noise of $\sigma_\epsilon=0.26$.
    
    \item CL+LoS+${\sigma_\epsilon}$: This mock accounts for both systematic uncertainty due to line-of-sight structure and statistical uncertainty due to shape noise. It includes structure along the full line-of-sight and shape noise of $\sigma_\epsilon=0.26$.
\end{itemize}
A central part of this paper is to quantify the impact of line-of-sight structure on cluster weak lensing studies with upcoming Euclid data. To do this accurately, we must forward-model the source redshift distribution expected for Euclid, as provided by \cite{Mellier_2025}. To this end, we discretise this source redshift distribution into bins corresponding to the redshifts used in our single source plane mocks, assuming a fixed number of galaxies per bin. We then randomly associate every source galaxy on our uniform grid with a source redshift bin and assign to it the reduced shear from the corresponding single source plane mock. In this way, we construct a Euclid-like mock from a series of single source plane mocks.
However, this procedure introduces stochasticity in our mocks, over which we later marginalise by repeated resampling of the random assignment of source redshift bins to our sources. In this manner, we construct two more Euclid-like mocks, which are representative for upcoming Euclid data:

\begin{itemize}
    \item Euclid+CL: This mock combines the CL+$\sigma_\epsilon$ series of single source plane mocks into a mock with a source redshift distribution expected for Euclid. It includes shape noise of $\sigma_\epsilon=0.26$, but no line-of-sight structure. Instead, it includes only a thin line-of-sight shell of $\pm 5$ Mpc around the cluster.

    \item Euclid+CL+LoS: This mock combines the CL+$\sigma_\epsilon$+LoS series of single source plane mocks into a mock with a source redshift distribution expected for Euclid. It includes shape noise of $\sigma_\epsilon=0.26$ and structure along the full line-of-sight, which varies now in length from source galaxy to source galaxy.
\end{itemize}

To ensure the validity of our weak lensing formalism, we mask all source galaxies with $\kappa>0.9$ and $|\vec{g}|>1$. This roughly follows the findings of \cite{Massey_2008}, who demonstrated that the weak lensing approximation remains accurate to within a sub-percent level up to $|\vec{g}|\approx0.93$.

The specifics of these mocks and their designators are summarized in Table \ref{tab:data}.

\section{Modelling}
\label{sec:fit}
In this section, we discuss how we measure the clusters, mass, concentration, axis ratio and BCG wobble from our mock weak lensing maps.

\subsection{Mass density profile}
\label{sec:models}
In this work, we model the mass density profile of our simulated clusters with the Navarro-Frenk-White (NFW) profile, an empirical model that accurately describes the mass density profiles of dark matter halos in N-body simulations \citep{Navarro_1996,Navarro_1997}. 
Since the influence of line-of-sight structure on the inferred density profile parameters will depend on the choice of model, we will consider both the spherically symmetric and the elliptical NFW mass density profile. The NFW profile has the asymptotic form of a broken power law and under the assumption of spherical symmetry (sphNFW) it takes the form of
\begin{equation}
    \rho_\text{NFW}(r) = \rho_\text{s}\cdot \Bigg(\frac{r}{r_\text{s}}\Bigg)^{-1}\Bigg(1+\frac{r}{r_\text{s}}\Bigg)^{-2},
\end{equation}
where $r$ is the radius, $\rho_\text{s}$ the characteristic density and  $r_\text{s}$ is the scale radius, which marks the transition from the $r^{-1}$ to $r^{-3}$ scaling of the density profile. The scale radius can be parametrised as a fraction $1/c_{200}$  of the halo's $R_{200}$, where $c_{200}=R_{200}/r_\text{s}$ is the concentration. With these definitions the spherical NFW profile can be parametrised with four parameters: two for the centre, the mass $M_{200}$, and the concentration $c_{200}$. 

Analytical expressions for the convergence and shear for the spherical NFW profile were written down for the first time in \cite{Bartelmann_1996}.
However, it is well established that dark matter halos are triaxial ellipsoidals (e.g. \citealp{Ying_2002,Kasun_2005,Allgood_2006}) and the assumption of spherical symmetry in lensing studies of galaxy clusters leads to biases of up to 40\% on the mass (e.g. \citealp{Feroz_2011, Bahe_2012,Herbonnet_2022, giocoli_2024}). It is therefore important for lensing studies to generalize the spherical NFW profile to an elliptical one (eNFW), and have expressions for the convergence and shear. Despite the apparent simplicity of the NFW profile, deriving the convergence and shear in the elliptical case has proven to be difficult.

Recently, the problem of convergence and shear for an eNFW profile has been solved analytically by \cite{Heyrovsky_2024}, by introducing ellipticity directly into the mass density profile. These analytical expressions give the exact convergence and shear emerging from an eNFW profile and at the same time allow for fast model evaluations needed for Bayesian parameter inference. 
We refer the reader to \cite{Heyrovsky_2024} for details on the models. However, we note that we have modified the convention with which ellipticity is introduced. Specifically, we ensure that the enclosed mass in an iso-surface density contour stays constant for a changing axis ratio, $q$. This means we introduce ellipticity by changing coordinates from radius $r$ to semi major axis, $a$, with the following coordinate transformation
\begin{equation}
    r(x,y) \rightarrow a(x,y) = \sqrt{x^2/q + q\,y^2}.
\end{equation}
Here $(x,y)$ are the Cartesian coordinates on the image,  $r(x,y)$ the magnitude of $(x,y)$ and $q$ the axis ratio of the projected mass density profile.

The eNFW model of \cite{Heyrovsky_2024} has 6 free parameters: ($x_\text{c}, y_\text{c}$), the Cartesian coordinates of the centre; $q$, the axis ratio of the projected mass density profile; $\phi$, the orientation of the major axis with respect to the x-axis; $a_\text{s}$, the scale semi-major-axis; and $\kappa_\text{s}$, the halo convergence parameter. The parameter $\kappa_s$ can be related to the characteristic density $\rho_\text{s}$ of the 3D density profile,  $a_\text{s}$, the critical density of the Universe, and a factor that depends on the 3D orientation of the ellipsoid w.r.t. the line-of-sight. See equation 22 in \cite{Heyrovsky_2024} for details on this. 
This model assumes a constant ellipticity for the dark matter halo, contrary to the demonstrated radial dependence of the ellipticity in simulated halos (e.g. \citealp{Allgood_2006, Schneider_2012}). We note that this is a simplifying assumption in this study, which can be improved upon in future work.

Following the same reasoning as in the spherical case, we can swap $a_\text{s}$ and $\kappa_\text{s}$ for $M_{200}$ and $c_{200}$. 

In this work, we employ both spherical and elliptical NFW models to measure mass and concentration and compare the results. In studies focusing on these parameters, the cluster centre is typically fixed by using a specific tracer. To maintain consistency with such studies and simplify our statistical analysis, we fix the centre at the position of the BCG (see Section \ref{sec:BCG} for details). This choice minimises biases from profile mis-centring, as collisionless dark matter simulations predict that the BCG’s offset from the potential centre is generally well below the softening length \citep{roche2024, Schaller_2015}. In observational studies, where the optimal tracer of the potential centre is uncertain, alternative choices include the X-ray centre, satellite distribution, or strong lensing data (e.g., \citealp{von_der_Linden_2014,Wang_2018,Oguri_2012}). 

Once the cluster's centre is assumed, a choice must be made for the radial fitting range in the weak lensing analysis. An inner fitting radius is used to minimise baryonic effects, while an outer radius helps mitigate the influence of large-scale structure. Previous studies have shown that the specific choice of these radii can introduce biases in weak lensing mass measurements (e.g., \citealp{Becker_2011, Bahe_2012, Lee_2018}). In this work, we do not set an outer fitting radius due to the limited field-of-view, which extends only to $R\sim 2\text{ Mpc}$. However, we mask all source galaxies within 30 arcsec of our assumed centre (the BCG) when inferring mass, concentration, and axis ratio. When applying the sphNFW model, we convert the relevant mock to reduced tangential shear. 

To measure the BCG wobble we fit the full six-parameter eNFW model to the mocks. To maximize the sensitivity, we do not mask out the central region of the cluster. In this approach, we effectively marginalise over all other parameters in the eNFW model.

A summary of the three different models that we use in this work is provided in Table \ref{tab:models}. 

Our Euclid-like mocks include a realistic distribution of source galaxy redshifts, for which we assume an effective source redshift in our modelling. Specifically, we solve equation \ref{eq:zeff} for $z_\text l=0.225$ and $n(z_\text s)$ from \cite{Mellier_2025}, finding $z_\text{eff}=0.60$. While this procedure matches the average lensing efficiency of the full source redshift distribution, it still approximates all source galaxies to lie on a single source plane at $z=z_\text{eff}$. This approximation can lead to additional bias in the inferred lensing signal due to the non-linear dependence of reduced shear on lensing efficiency. As a result, best-fit parameters (most notably mass) can receive additional bias from this simplification. 
Naturally, for the single source plane mocks, we assume perfect knowledge of the source plane redshift and take this value as input for our modelling.
\begin{table}
\caption{Overview NFW models}
\label{tab:models}
\centering

\begin{tabular}{lc c c c c}
\hline\hline
Model& $i_{c}$& $M_{200}\,/\,10^{14}\text{ M}_\odot$ & $\sigma_\text{c}$& $e_i$\\
\hline
sphNFW             &  $\,\,i_\text{BCG}$& [$0.5,100$]& [-3,3]&0\\
eNFW$_4$&  $\,\,i_\text{BCG}$& [$0.5,100$]& [-3,3]& [-1,1]\\
eNFW$_6$& [$i_\text{BCG} \pm$ 25"]& [$0.5,100$]& [-3,3]& [-1,1]\\
\hline
\end{tabular}
\tablefoot{
The three different models we use in this work: i) ``sphNFW'', a spherically symmetric NFW model with centre fixed at the BCG; ii) ``eNFW$_4$'', an elliptical NFW model with the centre fixed at the BCG; and iii) ``eNFW$_6$'', the full 6-parameter elliptical NFW model. Entries without square brackets denote fixed parameters, while those in square brackets represent the prior limits of a free parameter. The prior on mass follows a log-uniform distribution, whereas all other parameters have uniform priors. In this table, $i$ denotes either the $x$ or the $y$ coordinate; $\sigma_\text{c}$ the number of standard deviations of the scatter above/below the \cite{Ludlow_2016} concentration-mass relation; and $e_i$ the decomposed ellipticity defined in equation \eqref{eq:decomp_e}.
}

\end{table}
\subsection{Bayesian parameter estimation}
We perform Bayesian parameter estimation with the fitting library of \texttt{PyAutoLens} \citep{Nightingale_2021}. This is an open-source Python package originally developed for strong gravitational lensing with various features, such as (strong) lens modelling, (strong) lens simulations, and ray tracing. In this work we make use of its elaborate library designed for lens modelling. We make use of the \texttt{NFWMCRScatterLudlow} mass density profile, which samples the concentration, $c_{200}$, as a number of  ``$\sigma_\text{c}$'' standard deviations (in units of dex) above or below the \cite{Ludlow_2016} concentration mass relation, such that 
\begin{equation}
    \text{log}_{10}(c_{200}) = \text{log}_{10}(c(M)) + 0.15 \sigma_\text{c}.
\end{equation}

Where $c(M)$ is the \cite{Ludlow_2016} concentration-mass relation and 0.15 is the standard deviation of the scatter in the concentration mass relation in units of dex \citep{Wang2020}.  

The axis ratio and orientation of the major axis are sampled at the level of the decomposed ellipticity $\vec{e} = (e_x,e_y)$, which satisfies
\begin{equation}
    q = \frac{1 - |\vec{e}|}{1 + |\vec{e}|},\,\,\,\, \,\, \phi = \text{arctan}(e_y/e_x).
    \label{eq:decomp_e}
\end{equation}
Within the framework of \texttt{PyAutoLens} we use \texttt{Nautilus} \citep{nautilus}, which is a Python package designed for Bayesian posterior and evidence estimation. It improves on traditional Markov chain Monte Carlo (MCMC) methods by combining importance nested sampling (INS) \citep{Feroz_2019} with neural networks. For our model fits, we use 1000 live points. This choice provides a satisfactory balance between accuracy, precision and computational efficiency, as determined through tests with our mock data sets.

We adopt uniform priors except for $M_{200}$, for which we adopt a log-uniform prior due to its wide dynamical range. This is common practice in cluster weak lensing studies (e.g., \citealp{Sereno_2013, Umetsu_2014, Umetsu_2020, Okabe_2019}) and planned for use in the Euclid cluster weak lensing program (e.g., \citealp{giocoli_2024, Sereno_2024, Ingoglia_2025}). Alternative choices, such as a uniform mass prior, can be made and would drive the inferred masses slightly upward. Due to computational constraints, we do not explore the interplay between the choice of mass prior and the effects of line-of-sight structure, and restrict our study to a log-uniform mass prior. We acknowledge that our results may change if a uniform mass prior is assumed instead.
The priors are defined within the following limits:
\begin{itemize}
    \item Mass density profile center: A square of width 50 arcsec (187 kpc at $z=0.225$) centred on the BCG. 
    \item Mass: A range between $5\times 10^{13}\text{ M}_\odot$ and $1\times 10^{16}\text{ M}_\odot$.
    \item Number of standard deviations above/below the \cite{Ludlow_2016} concentration mass relation: 
A range of $\pm 3$ standard deviations. 
    \item Decomposed ellipticity: The full range of physically meaningful values: $-1<e_{x,y}<1$. 
\end{itemize}

A summary of these priors is listed in Table \ref{tab:models}.

In this study, we aim to quantify the impact of line-of-sight structure on weak lensing observables of galaxy clusters, in case no attempts are made to correct for them. In principle, one can mitigate its impact on the inferred precision by including non-diagonal terms in the covariance matrix, which are given by the two-point shear correlation function (see e.g. \citealp{Oguri_2010}). However, we keep our covariance matrix diagonal and quantify the impact of line-of-sight structure on the scatter. We note that including off-diagonal elements in the covariance matrix will not mitigate any biases induced by line-of-sight structure that we present in this paper.

\subsection{Brightest cluster galaxy}\label{sec:BCG}
One aspect of this paper is to study the impact of line-of-sight structure on the measured  BCG wobble. These measurements rely on identification of two points: i) the bottom of the gravitational potential, measured as the centre of the total mass density profile obtained through gravitational lensing  and ii) the BCG centre, determined observationally from the peak of the stellar light distribution.

To determine the centre of the BCG, we generate surface density maps using stellar particles in the simulation. We do this by taking a thin shell of $\pm 5$ Mpc around the cluster in the lightcone particle data and projecting all the particles on a 2 dimensional grid. This grid has a physical field-of-view of 1 Mpc and a resolution equal to the softening length, which we have chosen so that the particle per pixel count is sufficiently high.
\begin{figure*}[t]
    \centering
    \includegraphics[width=1\linewidth]{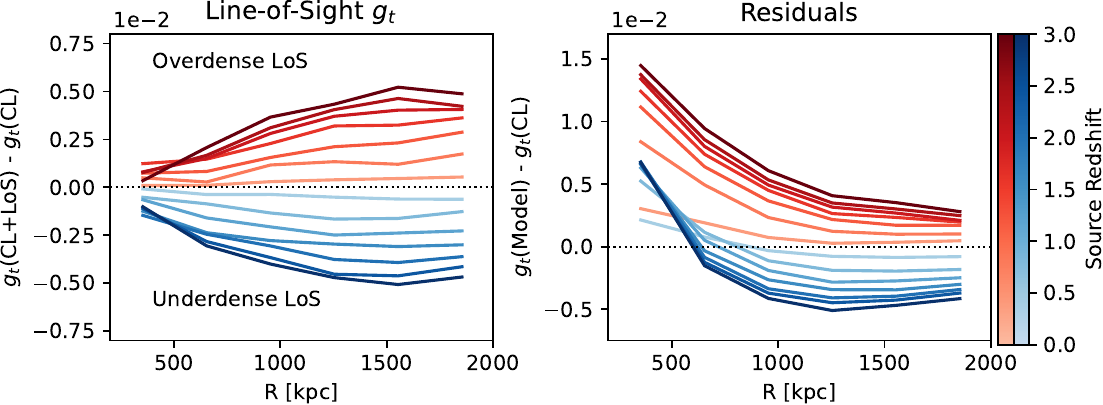}
    \caption{The sample is subdivided into two subsamples having an over-dense (red) or under-dense (blue) line-of-sight for a given source redshift (shaded colours). Left: Median azimuthally averaged reduced tangential shear contribution from the line-of-sight, calculated as the difference between the reduced tangential shear from the CL+LoS mock and the CL mock, as a function of radius. Right: Median azimuthally averaged residuals, calculated as the difference between the reduced tangential shear from the best-fit model and the CL mock. On average, the residuals for clusters lying along over-dense and under-dense sight-lines are asymmetrically distributed around zero, potentially leading to a bias.}
    \label{fig:residuals}
\end{figure*}
We assume that the stellar current particle masses trace the stellar light and use the resulting stellar surface density map as input for the peak-finding algorithm \texttt{SExtractor} \citep{Bertin_96}. To improve the accuracy of the peak identification, we up-sample the input maps through linear interpolation by a factor of 10 and apply a Gaussian filter with a standard deviation of 20 pixels. 
In the \texttt{SExtractor} output, we identified the source with the highest ``flux'' (\texttt{MAG\_ISO}) as the BCG and retrieved its centroid with the \texttt{X\_IMAGE} and \texttt{Y\_IMAGE} outputs.  

\section{Results}
\label{sec:results}
We study the impact of line-of-sight structure on weak lensing analyses of clusters, assuming the NFW profile variations outlined in Table \ref{tab:models}. In the following, we compare the results of mass density profile reconstructions across the six mocks described in Table \ref{tab:data}. For each single source plane mock, we perform a Bayesian parameter inference for mock setups with source plane redshifts of 0.8, 1.2, 1.6, 2.0, 2.4 and 3.0. Source redshifts below this range result in non-detections for low-mass clusters for mocks with non-zero shape noise, complicating fair comparisons between the samples. For the CL mock, the results are independent of source redshift due to the absence of both shape noise and line-of-sight structure. Therefore, we perform the parameter inference on the CL mock only for a source plane redshift of 0.8 and use these results for all other source redshifts as well.

\subsection{Residuals sphNFW fit on CL+LoS mock}
\label{sec:Residuals}
To illustrate the effect of line-of-sight structure on weak lensing observables, we first consider a simple case: fitting the sphNFW model to our CL+LoS mock. For each source redshift, we subdivide the sample into two groups based on whether the cluster lies along an over-dense or under-dense line-of-sight.
We assume that clusters along over-dense sight-lines have overestimated masses, while those along under-dense sight-lines have underestimated masses. Using mass bias as an estimator, we classify clusters with a mass bias below the sample median as “under-dense LoS” and the rest as “over-dense LoS.”

To evaluate this subdivision, we examine the azimuthally averaged reduced tangential shear contribution from the line-of-sight, hereafter $\bar{g}_\text{t, LoS}(R)$. To compute this quantity, we subtract the CL mock from the CL+LoS mock, and average the result azimuthally in nine linear bins between $0.5$ and $9$ arcminutes.

The left panel of Figure \ref{fig:residuals} shows the median $\bar{g}_\text{t, LoS}(R)$ profiles accross our subsamples and source redshifts. Angular bins are converted to physical distances using the mean cluster redshift of $z = 0.225$. under-dense and over-dense LoS classifications are plotted in blue and red respectively, with varying shades representing different source redshifts.
\begin{figure*}[t]
    \centering
    \includegraphics[width=0.85\linewidth]{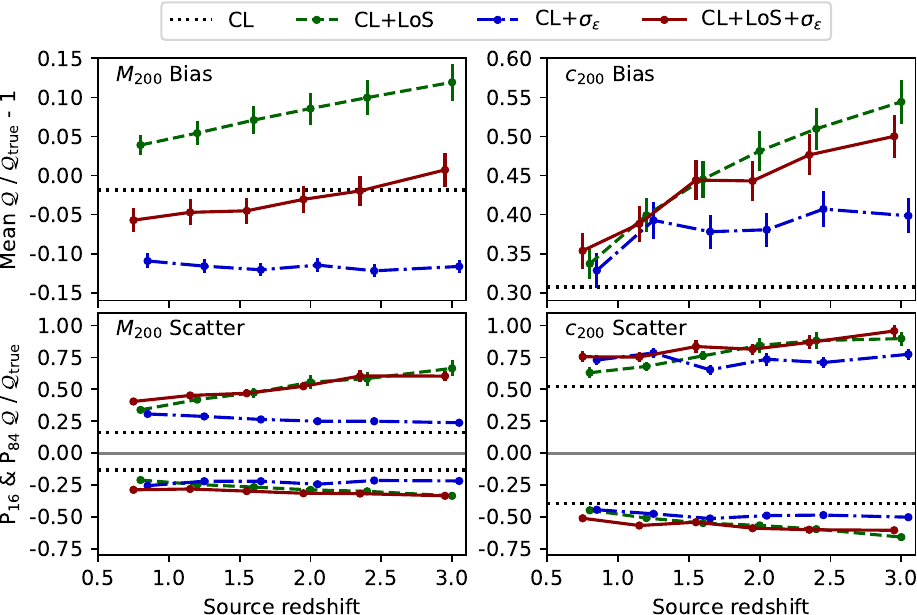}
    \caption{Applying the sphNFW model we show the mean bias (top row) and scatter (bottom row) as a function of source redshift for the cluster's mass (left column) and concentration (right column), for our four mocks CL (black/dotted), CL+LoS (green/dashed), CL+$\sigma_\epsilon$ (blue/dot-dashed) and CL+LoS+$\sigma_\epsilon$ (red/solid). We estimate the bias of quantity ``$\mathcal{Q}$'' as the mean of the relative error distribution minus 1. We estimate the the upper bound scatter ($+$) and lower bound scatter ($-$) using the 84th- and 16th percentile of the relative error distribution, respectively.}
    \label{fig:sph_mc_mean}
\end{figure*}
Our subdivision shows how over-dense and under-dense sight-lines are symmetrically distributed around zero. In other words, the median $\bar{g}_\text{t, LoS}(R)$ profile for the full sample is zero. This is expected, since an equal number of sight-lines pass through over-dense and under-dense regions of the Universe, causing their contributions to the median reduced tangential shear profile to average out to zero. 
The absolute value of the median $\bar{g}_\text{t, LoS}(R)$ profile increases with radius because the annular area grows, leading to greater variance in the projected mass density along the line-of-sight. The absolute value of the median $\bar{g}_\text{t, LoS}(R)$ profile increases with source redshift as well, due to the longer line-of-sight and boosted lensing efficiency.

In the right panel of Figure \ref{fig:residuals}, we plot the median azimuthally averaged residuals as a function of radius for the under- and over-dense LoS subsamples. The residual is computed by subtracting the CL mock from the best-fit model.
We find that over-dense LoS clusters consistently have overestimated profiles, with residuals increasing toward the centre. In contrast, under-dense LoS clusters have underestimated profiles at large radii but overestimated profiles in the inner regions.

Although the reduced tangential shear profile itself remains unbiased on average, the residuals of the under-dense and over-dense LoS subsamples are asymmetric around zero, potentially introducing biases in the inferred best-fit parameters. Physically, this suggests that the sphNFW profile provides a better fit for clusters along under-dense sight-lines than for those along over-dense sight-lines.

\subsection{Mass, concentration and axis ratio}
In the previous section, we demonstrated that in the limit of infinite signal-to-noise, the sphNFW model provides a better fit for clusters along under-dense sight-lines than for those along over-dense ones. In this section, we quantify the resulting biases in the inferred cluster mass, concentration and axis ratio. We also examine the scatter and the inferred precision of these parameters in relation to the effects of line-of-sight structure.

We begin by considering our single source plane mocks to study the interplay between shape noise and line-of-sight structure, and how this depends on source redshift. We first assess the impact on the best-fit parameters of the sphNFW model in Section \ref{sec:sph_nfw}, after which we discuss the same for the eNFW$_4$ model in Section \ref{sec:enfw}. Subsequently, we turn to our Euclid-like mocks to quantify the expected biases in cluster weak lensing analyses with Euclid data in Section \ref{sec:nz}. Finally, we discuss the estimated uncertainties on the free parameters of both models in Section \ref{sec:estimated_uncertainty}.

\subsubsection{Spherical NFW model}
\label{sec:sph_nfw}
The sphNFW model has two free parameters: mass and concentration.
As a ground truth for the mass and concentration we use the SOAP catalogued $M_{200}$ and $c_{200}$ from the simulation's post-processing files\footnote{SOAP was run on the simulation snapshots, cataloguing cluster properties at either $z=0.20$ or $z=0.25$, depending on which is closer. Considering the elapsed time between $z=0.225$ and $z=0.200$ (approximately equal to the time between $z=0.250$ and $z=0.225$), and using the average halo accretion rate of \cite{Correa_2015}, we estimate that the halo mass changes by $\sim$ 2\%. Since the sample's redshift distribution is uniform, this will not lead to a bias, since half of the clusters will have slightly overestimated masses, while the other half will be slightly underestimated. This effect contributes minimally to the scatter, with a maximum impact of $\sim$2\%, rendering it subdominant. Given the small change in mass, we expect the change in concentration to be subdominant as well.}. SOAP calculates $c_\text{200}$ using a slightly modified version of the method described by \cite{Wang_2023}, which estimates concentration based on the first-order moment of the mass density distribution, $R_\text{1}$. This method uses a fifth-order polynomial fit to the $R_\text{1}$-concentration relation for concentrations ranging between 1 and 1000.
\begin{figure*}[ht]
    \centering
    \includegraphics[width=1\linewidth]{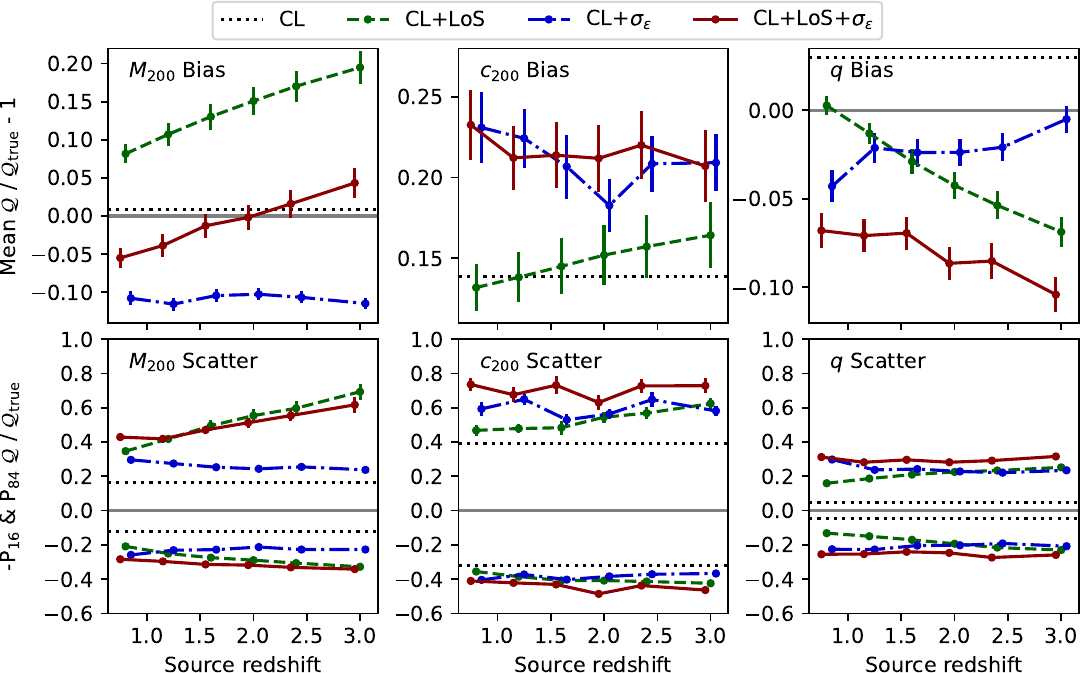}
    \caption{Same as Figure \ref{fig:sph_mc_mean}, but now employing the eNFW$_4$ model, which treats the axis ratio ($q$) as a free parameter (right column).}
    \label{fig:enfw4_mcq}
\end{figure*}
In Figure \ref{fig:sph_mc_mean} we show the mean\footnote{We show the median bias in Figure \ref{fig:sph_mc_median} in Appendix \ref{appendix:median_biases}.} bias (top row) and scatter (bottom row) as a function of source redshift for the mass (left column) and concentration (right column) for our four mocks:  CL+LoS (green/dashed), CL+$\sigma_\epsilon$ (blue/dot-dashed), CL+LoS+$\sigma_\epsilon$ (red/solid) and CL (black/dotted).
We use the mean of the relative error distribution minus 1 as our estimator for the bias, while for scatter, we use the 16th and 84th percentiles of the relative error distributions, which are plotted as negative and positive values, respectively. Plotting the scatter in this manner allows us to observe any asymmetry induced in the relative error distributions. The error bars indicate the bootstrapped 1$\sigma$ uncertainty on the mean or percentile.

In the upper panels of Figure \ref{fig:sph_mc_mean}, we observe that in our CL mock the sphNFW model intrinsically underestimates the mass and overestimates the concentration. This bias arises from projection effects: When a halo is observed along its major axis, the mass is overestimated, whereas projections along the other two axes underestimate it (e.g., \citealp{Oguri_2011,Bahe_2012, giocoli_2024}), leading to a net negative bias on average. The concentration compensates for this through the concentration-mass degeneracy, leading to an overestimated concentration. We refer the reader to e.g. \cite{Oguri_2011} for a more detailed discussion on this. It has been shown that this weak lensing mass bias can be reduced by fixing the concentration, either to a fixed value or via a theoretical concentration-mass relation (e.g., \citealp{Lee_2018, giocoli_2024}). Furthermore, we observe that the intrinsic scatter in the mass is around a factor of two smaller than that in the concentration. This is a well-known result and can again be explained by projection effects.

The results from the CL+LoS mock in the bottom panels indicate that the intrinsic effect of line-of-sight structure is to generate a positive skew in the relative error distributions for both mass and concentration. As a result, the mean biases in mass and concentration increase with source redshift.

This can be understood in context of the previous section, where we saw that the sphNFW model provides a better fit to clusters along under-dense sight-lines than those lying along over-dense ones. This leads to a model bias, whose quantitative effect on the mass and concentration we observe here. From the right panel of Figure \ref{fig:residuals} it is evident that the best-fit model is indeed overly concentrated on average.

With regards to the CL+$\sigma_\epsilon$ mock, we observe that the scatter decreases slightly due to the boosted lensing efficiency between $z_\text{s}=0.8$ and $z_\text{s}=3.0$. Since there is no skew, we observe that both the mean mass and concentration biases do not vary with source redshift. We argue that the $z_\text s = 0.8$ point for the mean concentration bias that breaks this trend is due to some low signal-to-noise clusters with poorly constrained concentrations.
 
Aside from the presence of line-of-sight structure, the key difference between these mocks is that the CL+LoS mock assumes an infinite signal-to-noise, whereas the CL+$\sigma_\epsilon$ mock has a realistic one. As a result, in the CL+LoS mock, all radii contribute statistically, whereas in the CL+$\sigma_\epsilon$ mock, the signal-to-noise drops below unity at a certain radius. Combined with the concentration-mass degeneracy, this leads to altered biases in mass and concentration.

We study the combined effects of realistic signal-to-noise and line-of-sight structure using the CL+LoS+$\sigma_\epsilon$ mock. We observe that the relative error distributions are skewed, similarly to the CL+LoS mock. Therefore, the mean mass and concentration biases follow a trend toward more positive biases as well. However, the effects of finite signal-to-noise changes its amplitude with respect to the CL+LoS mock. 
A comparison between the CL+$\sigma_\epsilon$ and CL+$\sigma_\epsilon$+LoS mocks shows that line-of-sight induced biases in both mass and concentration persist under the assumption of realistic signal-to-noise. 
Further, we note that for $z_\text s \gtrsim1.2$, line-of-sight structure dominates the upward scatter in the mass estimates. For the scatter in concentration the effects of line-of-sight structure are less pronounced, indicating that other factors are likely more influential.

\begin{figure*}[ht]
    \centering
    \includegraphics[width=0.95\linewidth]{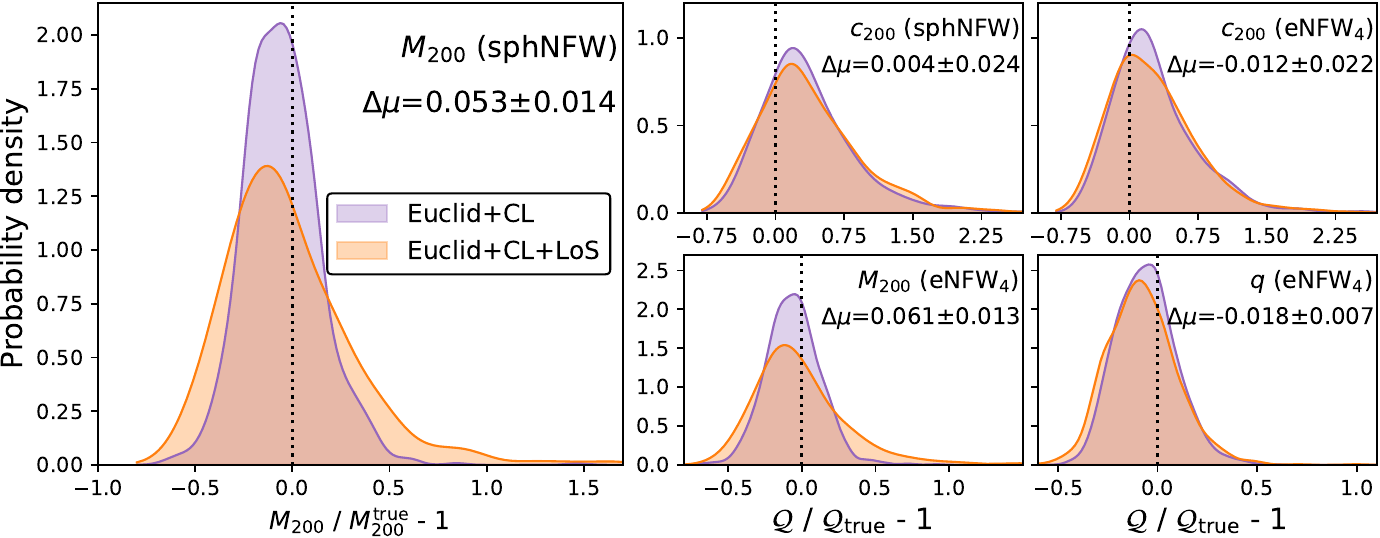}
    \caption{Relative error distributions for the best-fit parameters in the Euclid+CL (purple) and Euclid+CL+LoS (orange) mocks. We show the relative error distribution for mass (left panel); concentration (right panel/top-left) under the assumption of the sphNFW model; and mass (right panel/bottom-left); concentration (right panel/top-right); and axis ratio (right panel/bottom-right) under the assumption of the eNFW$_4$ model. We report the difference of the means ($\Delta \mu$) and it's bootstrapped uncertainty on the top right of each panel. For Euclid-like data, line-of-sight structure positively biases mass estimates with the sphNFW model on the level of $+5.3\pm1.4,$\%, which is significant at 3.5$\sigma$.}
    \label{fig:Euclid_mocks_mcq}
\end{figure*}

\subsubsection{Elliptical NFW model}
\label{sec:enfw}
This subsection discusses the results for the eNFW$_4$ model, which adds the axis ratio of the projected mass density profile as an additional degree of freedom. As a ground truth for the axis ratio, we use the axis ratio of the best-fit eNFW$_4$ model on 2 arcsec resolution convergence maps.
Similarly to Figure \ref{fig:sph_mc_mean}, Figure \ref{fig:enfw4_mcq} shows the mean\footnote{We show the median bias in Figure \ref{fig:enfw4_mcq_median} of Appendix \ref{appendix:median_biases}.} bias and scatter for mass, concentration and axis ratio as a function of source redshift across our four mocks. 

First, we discuss the results on the mass and concentration. The CL mock demonstrates that, compared to the sphNFW model, the mean concentration bias decreases by  $\sim$20\% point, while the absolute median mass bias shows a slight reduction. This suggests that projection effects are partially mitigated by the added flexibility of the eNFW$_4$ model.

For the CL+LoS mock, we observe trends that closely mirror those found for the sphNFW model. However, in this case, we find that the skew in the relative error distribution is even stronger, especially in the mass. Similarly, the CL+$\sigma_\epsilon$ mock shows trends consistent with the sphNFW model, with both the mean mass and concentration biases roughly constant with source redshift. Comparing the mean mass bias in the CL+$\sigma_\epsilon$ and CL+LoS+$\sigma_\epsilon$ mocks, we observe that line-of-sight structure induces additional bias, consistent with our findings for the sphNFW model.
In contrast, for the concentration, we now see no significant bias resulting from the effects of line-of-sight structure. This highlights that the biases associated with line-of-sight structure can be model dependent.

Regarding the scatter, we observe an overall decrease compared to the sphNFW model, which again relates back to the partial modelling of halo triaxiality. As with the sphNFW model, line-of-sight structure dominates the positive scatter in mass for $z_\text s \gtrsim1.2$, while its effect on concentration scatter is weaker, likely due to other dominant factors.

Moving on to the right most column, we observe for the axis ratio that line-of-sight structure biases this measurement low, meaning clusters appear more elliptical than they truly are. This can be understood, as halos along the line-of-sight, projected in the cluster's outskirts, will elongate the best fit mass model in order to account for their signal. On the other hand, the line-of-sight contributes minimally to the scatter in the measured axis ratio.

\begin{figure*}[t]
    \centering
    \includegraphics[width=\linewidth]{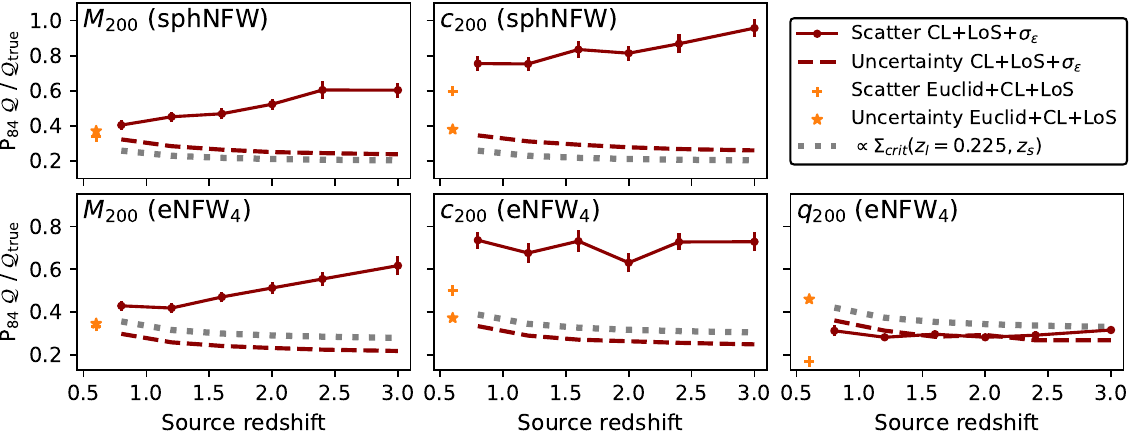}
    \caption{Comparison of the estimated uncertainty (\texttt{Nautilus}, diagonal covariance matrix) and true scatter (84th percentile relative error distribution). The top panels show the results for the free parameters of the sphNFW model, while the bottom panels show the same for the eNFW$_4$ model. For the CL+LoS+$\sigma_\epsilon$ mock, the estimated uncertainty and true scatter are shown as red/dashed and red/solid lines, respectively. For the Euclid+CL+LoS mock, the estimated uncertainty and true scatter are shown in orange stars and crosses, respectively. A gray/dotted curve indicates a quantity inversely proportional to the lensing efficiency. Under assumption of a diagonal covariance matrix, line-of-sight induced scatter in mass estimates is not accounted for in the Bayesian parameter inference.
    }
    \label{fig:estimated_uncertainty}
\end{figure*}

\subsubsection{Euclid-like mock data}
\label{sec:nz}
In the previous sections we have seen that line-of-sight structure can lead to additional bias in parameter inference, with the magnitude of this bias increasing with source redshift. A key simplification in the mocks we considered is the assumption of single source planes. Next, we want to take into account the source redshift distribution expected for Euclid, to quantify the significance of line-of-sight induced biases for upcoming Euclid data. Toward this end, we study our Euclid-like mocks (Euclid+CL and Euclid+CL+LoS), which have sources randomly sampled from the expected source redshift distribution for Euclid, as provided by \citealp{Mellier_2025} (see Section \ref{sec: mocks} for details). In the modelling of the cluster's lensing signal, we assume an effective source redshift of $z_\text{eff}=0.6$ (see the end of Section \ref{sec:models} for details). 

By randomly sampling the redshift of the sources on our uniform grid, we introduce stochasticity in our analysis. We marginalise over this stochasticity by rerunning our Bayesian inference 10 times for each cluster, each time resampling the source redshifts. We then report the mean best-fit parameters over these 10 inferences for the remainder of our analysis. We have verified that increasing this number to 20 times does not significantly affect our results in fitting the sphNFW model to the Euclid+CL+LoS mock.

In Figure \ref{fig:Euclid_mocks_mcq} we plot the resulting relative error distributions of our model free parameters for the Euclid+CL mock in purple and the Euclid+CL+LoS mock in orange. In the large panel on the left, we show the relative error distribution for the mass, estimated by assuming the sphNFW model, which is a common approach in cluster weak lensing studies. In the right panels, we present — from left to right and top to bottom — the relative error distributions for: the concentration estimated with the sphNFW model, followed by the concentration, mass, and axis ratio estimated with the eNFW$_4$ model. In all panels, we show zero (black/dotted) and the difference in mean bias, $\Delta \mu$, with the bootstrapped 1$\sigma$ uncertainty.

As we discussed in previous sections, we observe in the left panel that line-of-sight structure induces a positive skew in the relative error distribution of the mass estimated with the sphNFW model. This results in a positive mean mass bias $+5.3\pm1.4 \%$, which is significant at $3.5\sigma$. When the mass is estimated with the eNFW$_4$ model, the mean mass bias associated with line-of-sight structure increases to $+6.1\pm1.3 \%$, which is significant at $4.7\sigma$.
In these mocks, the concentration is not biased significantly by line-of-sight structure. As seen before, the scatter in concentration is much larger than in the mass, rendering the effects of line-of-sight structure subdominant. Moving on to the axis ratio, we observe that line-of-sight structure biases these measurements low by $-2.0\pm0.7\%$ on average, which is significant at 2.9$\sigma$. As expected, the results for these mocks are roughly consistent with what we observed for comparing the single source plane mocks CL+$\sigma_\epsilon$ and CL+LoS+$\sigma_\epsilon$ at $z_\text s=0.8$. These results demonstrate that, for cluster weak lensing analyses with Euclid data, the impact of line-of-sight structure must be incorporated into the error budget to avoid significant biases in cluster mass and shape estimates.

\subsubsection{Inferred precision}
\label{sec:estimated_uncertainty}
Having established that, at least for the mass, line-of-sight structure contributes significantly to the scatter, we need to investigate whether this is accounted for in the estimated uncertainty as derived during the fit by \texttt{Nautilus}. Note that we have not included the effects of line-of-sight structure in the covariance matrix, which would help estimate the true uncertainty. To do this, we compute the mean 84th percentile of the Bayesian-inferred relative uncertainty from the CL+LoS+$\sigma_\epsilon$ and Euclid+CL+LoS mocks. The results are shown in Figure \ref{fig:estimated_uncertainty}, where the CL+LoS+$\sigma_\epsilon$ mock is represented by red/dashed lines, and the Euclid+CL+LoS mock by orange stars. We compare this estimated uncertainty to the 84th percentile of the true relative error distribution, represented by red/dashed lines and orange crosses for CL+LoS+$\sigma_\epsilon$ and Euclid+CL+LoS mocks, respectively. 
The top panels show this comparison for the mass and concentration estimated with the sphNFW model, while the bottom panels show this comparison for the mass, concentration and axis ratio estimated with the eNFW$_4$ model. Additionally, we plot in gray/dotted a curve proportional to the critical surface density, which is inversely proportional to the lensing efficiency at fixed lens redshift.

For the mass estimated by both models, we observe that the true scatter increases with source redshift, driven by the line-of-sight structure (as we discussed in Sections \ref{sec:sph_nfw} and \ref{sec:enfw}). In contrast, the estimated uncertainty decreases with source redshift, inversely proportional to the lensing efficiency. This indicates that, under the assumption of a diagonal covariance matrix, line-of-sight-induced uncertainties in mass estimates are not taken into account in the posterior distributions. The Euclid+CL+LoS mock shows that this effect is subdominant for upcoming Euclid data. However, for future, deeper, weak lensing surveys (e.g. the Nancy Grace Roman Space Telescope \citep{spergel2015, akeson2019}, scheduled for launch in 2027) this effect needs to be taken into account to avoid overestimating the precision on the mass. 

Regarding the concentration, we observe a large overestimation of precision for both mocks and models. However, as discussed in Sections \ref{sec:sph_nfw} and \ref{sec:enfw}, the large scatter in concentration is not driven by line-of-sight structure, but by other systematics such as projection effects. Therefore, this overestimation of precision in concentration is an issue separate from line-of-sight structure that needs to be addressed in future work. Regarding the axis ratio, we observe only small differences between true scatter and estimated uncertainty.

\subsection{Brightest cluster galaxy wobble}
\begin{figure}[t]
    \centering
    \includegraphics[width=\linewidth]{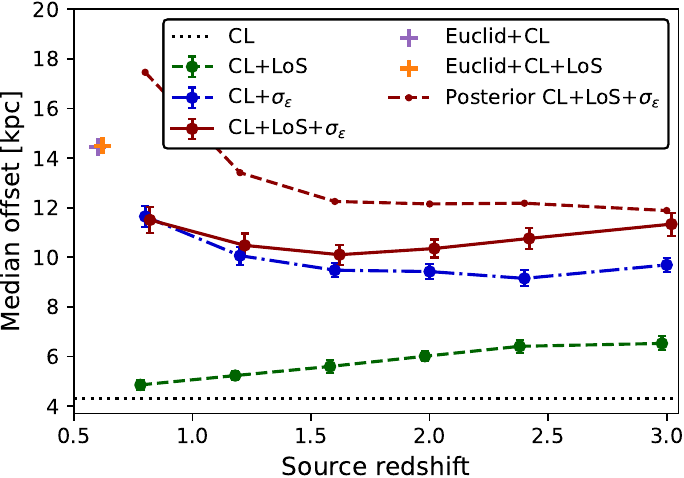}
    \caption{Median offset between the centre of the gravitational potential and the BCG (BCG wobble) as a function of source redshift for: the CL (black/dotted), CL+LoS (green/dashed), CL+$\sigma_\epsilon$ (blue/dot-dashed), CL+LoS+$\sigma_\epsilon$ (red/solid), Euclid+CL (purple/cross) and Euclid+CL+LoS (orange/cross) mocks. The error bars indicate the 1$\sigma$ bootstrapped uncertainty. The red/dashed line indicates the median wobble as expected from random sampling the posterior distribution for the centre on the CL+LoS+$\sigma_\epsilon$ mock. For upcoming Euclid data, measured median offsets of $\sim14$ kpc are expected for cold dark matter, with no significant contribution of line-of-sight structure.}
    \label{fig:fig_median_offset}
\end{figure}

Structure along the line-of-sight can induce mis-centring, biasing the measured offsets between the bottom of the gravitational potential and the brightest cluster galaxy (BCG), known as the BCG wobble. This subsection is dedicated to quantifying this effect relative to the statistical uncertainties introduced by shape noise.
As discussed in Section \ref{sec:fit}, we infer the bottom of the gravitational potential by applying the eNFW$_6$ model to the relevant mock. We plot the median BCG wobble measured in our samples as a function of  source redshift in Figure \ref{fig:fig_median_offset}. We show the results for the CL, CL+LoS, CL+$\sigma_\epsilon$ and CL+LoS+$\sigma_\epsilon$ single source plane mocks in the black/dotted, red/dashed, blue/dot-dashed and purple/solid curves, respectively. Additionally, we plot the results for the Euclid+CL and Euclid+CL+LoS mocks in purple and orange crosses, respectively.

We first analyse the CL and CL+LoS mocks to assess the intrinsic impact of line-of-sight structure on BCG wobble measurements. For the CL mock, we find a median BCG wobble of 4 kpc. Given that the high-resolution FLAMINGO run has a softening length of 2.85 kpc (for $z<2.91$), and that cold dark matter simulations predict offsets between the dark matter centroid and BCG to be smaller than the simulation’s softening length \citep{roche2024, Schaller_2015}, this result may seem unexpected. However, the lower offsets found by these studies can be attributed to their assumption of an idealised scenario in which the cluster’s mass distribution is perfectly known. In contrast, our result stems from biases in the eNFW$_6$ model, which arise from assumptions such as axisymmetry, as well as the observational limitation of a finite number of source galaxies. 
The CL+LoS mock demonstrates that the median BCG wobble increases with source plane redshift. This trend can be attributed to line-of-sight induced mis-centring of the mass density profile, indicating that in the limit of infinite signal-to-noise, line-of-sight structure biases the BCG wobble measurements.

Next, we measure the BCG wobble in the CL+$\sigma_\epsilon$ and CL+LoS+$\sigma_\epsilon$ mocks to study the impact of line-of-sight structure on measurements with a realistic signal-to-noise.
It is evident that introducing shape noise causes the median offset to increase substantially, due to the increased statistical uncertainty in the mocks. In the CL+$\sigma_\epsilon$ mock, we observe that the measured BCG wobbles decrease as a function of source redshift. This trend can be attributed to the increasing lensing efficiency at higher redshifts, which boosts the signal-to-noise.
For the CL+LoS+$\sigma_\epsilon$ mock, we find that the median offset initially decreases due to increased lensing efficiency, but then rises again due to the larger number of line-of-sight structures at higher source redshifts. The Euclid+CL and Euclid+CL+LoS mocks show, however, that the bias associated with line-of-sight structure can be completely neglected for upcoming Euclid data. We note that the median offsets for these mocks are unexpectedly high considering the $z_s=0.8$ single source plane mocks. This is due to poor sampling of the redshift distribution of source galaxies at small radii, where the offset measurements are most sensitive to. By marginalising over this effect in our mocks we introduce an additional source of statistical error, which further increases the measured median offsets.

Next, we investigate whether the line-of-sight induced uncertainty in the CL+LoS+$\sigma_\epsilon$+LoS mock is properly accounted for in the estimated posterior distributions of the Bayesian parameter inference. To this end, we computed the median wobble under the assumption that the bottom of potential and BCG coincide. In which case, the median offset depends solely on the uncertainty in the mass density profile centre. 
The magnitude of a two-dimensional vector with normally distributed components follows a Rayleigh distribution, whose median is given by $\sqrt{2ln(2)}\sigma_\text{comp}$, where $\sigma_\text{comp}$ is the standard deviation of the normally distributed vector components. Assuming that the posterior distributions for both components of the centre are normally distributed and, on average, identical, we can estimate the median wobble using this expression. The result is shown as the dashed red line in Figure \ref{fig:fig_median_offset}. This shows that the uncertainty estimate on the BCG wobble tends towards underestimating the true uncertainty. This indicates that any detection of a significant BCG wobble with weak lensing will be robust.

\section{Conclusions}
\label{sec:conclusions}
We quantify the impact of line-of-sight structure on the recovery of mass density profile parameters from weak lensing studies of galaxy clusters. Using the highest-resolution FLAMINGO simulation (L1\_m8; \citealp{Schaye_2023}), we construct a sample of 967 clusters with $M_{200}>3\times 10^{14} \text{M}_\odot$ and $0.20<z<0.25$. We generate several mock weak lensing maps, with a source galaxy number density of 30 galaxies/arcmin$^2$, consistent with expectations for the Euclid survey. 

To study the interplay between line-of-sight structure and shape noise, and how this evolves with source redshift, we construct four single source plane mocks. Additionally, we assess the significance of line-of-sight effects for upcoming Euclid data by constructing two more mocks with a realistic Euclid-like source redshift distribution (see Table \ref{tab:data}).
We perform Bayesian parameter inference with \texttt{Nautilus} \citep{nautilus} using three versions of the Navarro-Frenk-White (NFW) mass density profile to recover mass density profile parameters. For determining the mass, concentration, and axis ratio of the projected mass distribution, we use both spherical and elliptical NFW models (sphNFW and eNFW$_4$; see Table \ref{tab:models}), fixing the centre at the position of the brightest cluster galaxy (BCG). To determine the bottom of the gravitational potential (and thereby measuring the BCG wobble) we apply a six-parameter elliptical NFW model (eNFW$_6$; see Table \ref{tab:models}).
From our analysis, we conclude the following:

\begin{itemize}
    \item In the limit of infinite signal-to-noise, we find that the median azimuthally averaged reduced tangential shear profile remains unbiased by line-of-sight structure, meaning that the average profile is unaffected by variations in the line-of-sight contribution (left panel of Figure \ref{fig:residuals}). However, residuals from fitting a spherical NFW (sphNFW) indicates that it provides a better fit for clusters along under-dense sight-lines than for those along over-dense ones (right panel of Figure \ref{fig:residuals}), introducing a model bias. Indeed, as seen in Figure \ref{fig:sph_mc_mean}, increasing the source plane distance extends the line of-sight, generating a positive skew in the relative error distribution, and a resulting positive bias in both mass and concentration. These trends persist under the assumption of a realistic signal-to-noise, although to a lesser extend for the concentration. Regarding the elliptical NFW model, the same trends hold for the mass, although the bias in concentration largely disappears, highlighting the model dependence of these biases (Figure \ref{fig:enfw4_mcq}).\\

    \item Comparing two mocks representative of upcoming Euclid data, we conclude that line-of-sight structure induces an additional mass bias of $5.3\pm1.4$\%, when adopting the commonly used spherical NFW model (see left panel Figure \ref{fig:Euclid_mocks_mcq}). This bias is statistically significant at the $3.5\sigma$ level, highlighting its importance for accurate weak lensing masses. This mean mass bias increases further to $6.1\pm1.3$\% under the assumption of an elliptical NFW model. The concentration however, remains unbiased by line-of-sight structure for both models with large scatter due to other contributing factors (right panels Figure \ref{fig:Euclid_mocks_mcq}).
    
    Our conclusions regarding the sphNFW model contrast with previous findings for the weak lensing mass bias by \cite{Hoekstra_2011} and \cite{Becker_2011}. However, we note that \cite{Hoekstra_2011} assumed that clusters follow a perfect sphNFW profile, which differs significantly from the simulated clusters in this study. Additionally, \cite{Becker_2011} considered a maximum line-of-sight length of only 400 $h^{-1}$ Mpc, which may have been too short for this bias to manifest clearly.\\

    \item For source redshifts, $z_\text s\gtrsim1.2$, the effects of line-of-sight structure dominate the upper bound scatter in mass (see bottom rows of Figures \ref{fig:sph_mc_mean} and \ref{fig:enfw4_mcq}). When a diagonal covariance matrix is assumed, \texttt{Nautilus} fails to capture this additional scatter, resulting in an overestimation of the inferred precision (see Figure \ref{fig:estimated_uncertainty}). Our mocks representative of upcoming Euclid data, show that this effect will be subdominant for cluster weak lensing studies with Euclid (see Figure \ref{fig:estimated_uncertainty}). However, for future surveys such as the Nancy Grace Roman Space Telescope \citep{spergel2015, akeson2019}, which will produce significantly deeper weak lensing data, this effect will have to be taken into account. For the scatter in concentration, we find that line-of-sight structure plays a subdominant role compared to other factors, such as projection effects.\\

    \item Structure along the line-of-sight leads to an underestimation of the projected mass density profile's axis ratio, without significantly increasing the scatter (see Figure \ref{fig:enfw4_mcq}). For mocks representative of upcoming Euclid data, this bias amounts to $-2.0\pm0.7$\%, which is significant at $2.9\sigma$ (see Figure \ref{fig:Euclid_mocks_mcq}). This bias should be considered in studies using weak lensing measurements of halo ellipticity as a probe for self-interacting dark matter. Neglecting it could result in an underestimation of the halo's true roundness and, consequently, a loss of constraining power in the self-interaction cross-section of dark matter.\\
    
    \item Structure along the line-of-sight can cause mis-centring of the mass density profile, potentially biasing the median measured offset of the BCG from the potential centre (BCG wobble). For our single source plane mocks, this bias is only significant for high source redshifts, $z_\text s\gtrsim2$. Our mocks with a realistic Euclid-like source redshift distribution show that the effects line-of-sight structure are negligible compared to other sources of (statistical) uncertainty.
    Additionally, the dark matter self-interaction cross-section has been constrained through observations of the offsets between galaxies and the centre of the gravitational potential in merging clusters (see e.g., \citealp{Randall_2008, Harvey_2015,Sirks_2024}). While we have not specifically analysed BCG offsets from the bottom of the potential in merging clusters, we argue that our conclusions are still relevant in these cases.\\

    \item We forward-modelled measurements of the median BCG wobble with Euclid weak lensing data. Our results suggest that a measured offset of $\sim 14$ kpc is still consistent with cold dark matter (see Figure \ref{fig:fig_median_offset}). In this measured offset, the leading contributing factor is shape noise, with a small contribution from the sampling of the source redshifts at small radii. Moreover, we have shown that the Bayesian-inferred posterior for the mass density profile centre, tends to underestimate the true precision. This means that any statistically significant detection of a BCG wobble with weak lensing will be robust.
\end{itemize}

It is important to note that our choice of spatial distribution of source galaxies is a simplification. In reality, source galaxies are clustered and correlated with the line-of-sight structure we consider in this paper. This higher-order effect is not considered in this work, but was quantified to be negligible compared to statistical noise in \cite{Hoekstra_2011}.
Additionally, improvements could be made on the eNFW model by allowing the axis ratio of the projected mass density profile to vary with radius, providing a more realistic description of dark matter halo shapes.

In summary, we have shown that line-of-sight structure can introduce significant systematic uncertainty in weak lensing studies of galaxy clusters. While the average lensing signal of clusters remains unbiased, the inferred parameters are still susceptible to biases. This occurs because the NFW models used in this study provide a better fit for clusters along under-dense sight-lines than those along over-dense ones. Most notably, we find that in Euclid-like mock data, line-of-sight structure induces significant biases in cluster mass and shape estimates. Looking ahead, we predict that for future missions such as the Nancy Grace Roman Space Telescope, line-of-sight structure will become the dominant source of scatter in cluster weak lensing mass estimates. Assuming a diagonal covariance matrix in such analyses will then lead to a significantly overestimation of the precision.

As demonstrated in this work, future weak lensing studies of galaxy clusters striving for a high level of precision and accuracy must test their methods against the effects of line-of-sight structure and include these effects into their error budgets. Therefore, we advocate for the use of cosmological simulations with lightcone output that can self-consistently model effects along the full line-of-sight, in the calibration of cluster weak lensing pipelines.

\begin{acknowledgements}
This work was supported by the Swiss State Secretariat for Education, Research and Innovation (SERI) under contract number 521107294. The authors would like to thank John Helly and Roi Kugel for their role in the FLAMINGO project, which made this study possible.
This work used the DiRAC@Durham facility managed by the Institute for
Computational Cosmology on behalf of the STFC DiRAC HPC Facility
(\url{www.dirac.ac.uk}). The equipment was funded by BEIS capital funding via
STFC capital grants ST/K00042X/1, ST/P002293/1, ST/R002371/1 and ST/S002502/1,
Durham University and STFC operations grant ST/R000832/1. DiRAC is part of the
National e-Infrastructure. JWN is supported by an STFC/UKRI Ernest Rutherford Fellowship, Project Reference: ST/X003086/1.
\end{acknowledgements}

\bibliographystyle{aa}
\bibliography{refs} 

\begin{appendix}
\onecolumn
\section{The effect of up-sampling low-resolution lightcone output }
\label{appendix:smoothing}
The high-resolution FLAMINGO run (L1\_m8; \citealp{Schaye_2023}) we use in this study is accompanied by particle lightcone data for $z<0.25$, while for $z>0.25$ the lightcone output is stored as 13 arcsecond resolution full-sky HEALPix maps.
We aim to generate mass maps of the cluster and line-of-sight structure at a resolution of 2 arcseconds. To this end, we employ linear interpolation to up-sample the HEALPix maps to a resolution of 2 arcsec. In this section, we quantify the extent to which this impacts our final results.

\begin{figure}[t]
    \centering
    \includegraphics[width=\textwidth]{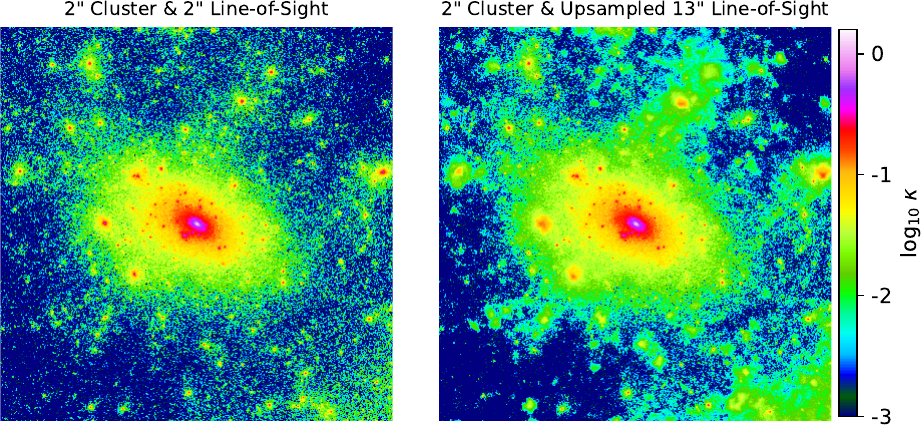}
    \caption{Convergence maps in log-scale of a cluster at $z=0.211$ with $M_\text{200}=3.32\times 10^{14}\text{ M}_\odot$, including line-of-sight structure up to $z = 0.25$. Left: Both cluster and line-of-sight structure mapped at 2 arcsecond resolution with the particle lightcone data. Right: Shell of the cluster from the particle lightcone data, with the other shells  from the HEALPix maps upsampled from 13 to 2 arcsecond resolution. In the outskirts of the cluster, we observe that up-sampling smooths the line-of-sight structure.}
    \label{app:LOS-convergence-map}
\end{figure}

In Figure \ref{app:LOS-convergence-map} we compare a convergence map of an example cluster with and without up-sampling the structure along the line-of-sight. Due to the restrictions on the particle lightcone data, we can only do this comparison for line-of-sight structure up to $z=0.25$. The left panel shows the cluster and line-of-sight structure ($z<0.25$) mapped at 2 arcsecond resolution with the particle lightcone data. The right panel shows the shell of the cluster mapped with the same particle lightcone data, while the other shells are up-sampled HEALPix maps. Qualitatively, one can see the smoothing effect of line-of-sight structure in the outskirts of the cluster.

To quantify whether the smoothing of line-of-sight structure impacts our final results, we forward-modelled both types of mocks (analogous to the CL+LoS mock in Section \ref{sec: mocks}), and fitted the eNFW$_6$ model (see Section \ref{sec:models}) to these mocks.
We find that in both mocks, $M_{200}$, $c_{200}$ and $q$ are all the same to within 0.1\%. Furthermore we found that the measured median BCG wobble is the same as well on the level of 0.1 kpc.

This test is limited to line-of-sight structure for $z<0.25$, and for longer sight-lines this effect will increase. However, the effect for $z<0.25$ is small and will decrease even further under addition of shape noise. Therefore, we conclude that up-sampling line-of-sight structure does not significantly affect our results.

\section{Median bias in mass, concentration and axis ratio}
\label{appendix:median_biases}

We have shown that line-of-sight structure induces a positive skew in the relative error distribution. Given this skew, showing the mean bias in our free parameters is the most representative of the effect of line-of-sight structure. However, in some cases, the median bias can be of interest as well.

\begin{figure*}[t]
    \centering
    \includegraphics[width=0.85\linewidth]{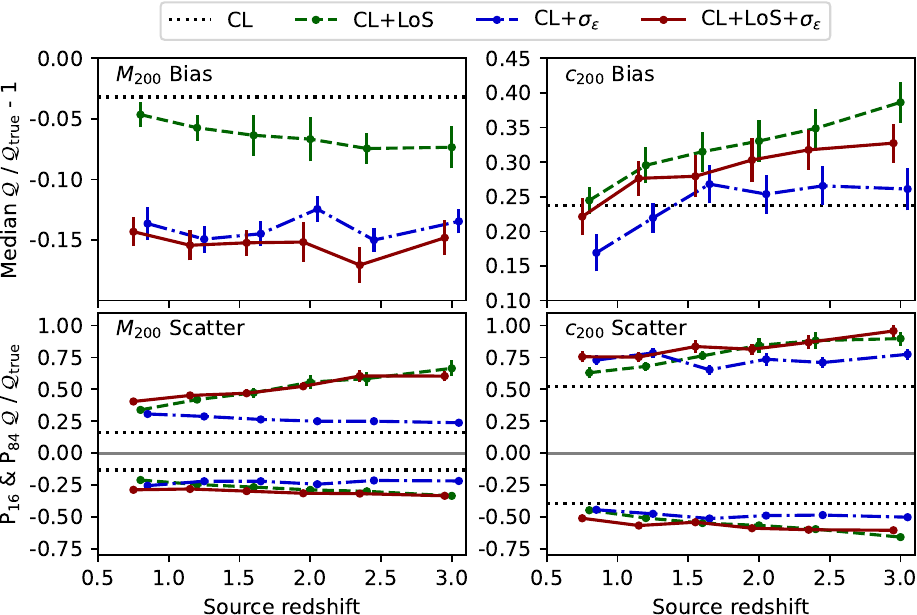}
    \caption{Same as Figure \ref{fig:sph_mc_mean}, but now with the median bias instead of the mean.}
    \label{fig:sph_mc_median}
\end{figure*}

\begin{figure*}[t]
    \centering
    \includegraphics[width=1\linewidth]{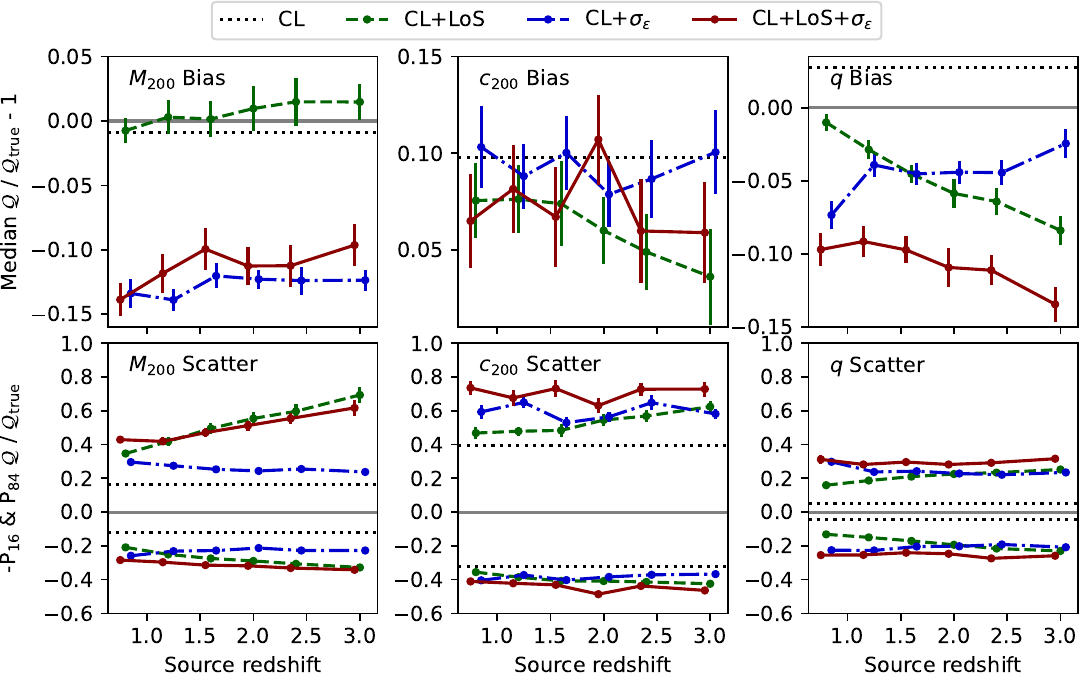}
    \caption{Same as Figure \ref{fig:enfw4_mcq}, but now with the median bias instead of the mean.}
    \label{fig:enfw4_mcq_median}
\end{figure*}

Therefore, we present our results for the median bias for the free parameters of the sphNFW and eNFW$_4$ models in Figures \ref{fig:sph_mc_median} and \ref{fig:enfw4_mcq_median}, respectively. The median is less sensitive to skewness than the mean. Consequently, the impact of line-of-sight structure on the median mass and concentration biases is less pronounced than on the mean bias.

\end{appendix}

\end{document}